\documentclass{IEEEcsmag}
\usepackage{amsmath,amssymb,amsfonts}
\usepackage{algorithmic}
\usepackage{algorithm}
\usepackage{array}
\usepackage[caption=false,font=normalsize,labelfont=sf,textfont=sf]{subfig}
\usepackage{textcomp}
\usepackage{stfloats}
\usepackage{verbatim}
\usepackage{graphicx}
\usepackage{cite}
\usepackage{bm}
\usepackage[utf8]{inputenc}
\usepackage{pifont}
\usepackage{upmath, colortbl}
\usepackage[dvipsnames]{xcolor}

\usepackage[hyphens]{url}
\usepackage[bookmarks=true,breaklinks=true,colorlinks,citecolor=black,linkcolor=black,urlcolor=blue]{hyperref}

\newcommand\circleOne{{\color{violet} \ding{202}}}
\newcommand\circleTwo{{\color{violet} \ding{203}}}
\newcommand\circleThree{{\color{violet} \ding{204}}}
\newcommand\circleFour{{\color{violet} \ding{205}}}
\newcommand\circleFive{{\color{violet} \ding{206}}}
\newcommand\circleSix{{\color{violet} \ding{207}}}
\newcommand\GreenCheck{{\color{ForestGreen} \ding{51}}}
\newcommand\RedX{{\color{BrickRed} \ding{55}}}
\newcommand\OrangeWarning{{\color{YellowOrange} \ding{115}}}

\def\mas{1}

\if\mas1
\def\GSterm{GMS}
\def\HSterm{HMS}
\def\ASterm{AMS}

\def\GStermshort{GMS}
\def\HStermshort{HMS}
\def\AStermshort{AMS}

\def\ASandHSTerm{AMS and the HMS}

\def\ASandGSandHSTerm{AMS, GMS, and HMS}

\else
\def\GSterm{Geometric Mean}
\def\HSterm{Harmonic Mean}
\def\ASterm{Arithmetic Mean}

\def\GStermshort{Gmean}
\def\HStermshort{Hmean}
\def\AStermshort{Amean}

\def\ASandHSTerm{arithmetic and harmonic means}

\def\ASandGSandHSTerm{arithmetic, geometric, and harmonic means}
\fi

\def\whatismean{aggregation metric}
\def\whatismeanpl{aggregation metrics}
\def\whatisipc{metric}
\def\whatisipcpl{metrics}
\def\whatisipccap{metric}
\def\whatisipccappl{metrics}

\def\RQterm{relevant \whatisipccap}
\def\RQtermPl{relevant \whatisipccappl}
\def\MQterm{measured \whatisipccap}

\def\RorMQterm{relevant or measured \whatisipccap}
\def\RandMQterm{relevant and measured \whatisipccap}
\def\RandMQtermpl{relevant and measured \whatisipccappl}

\jvol{59}
\jnum{2}
\paper{8}
\jmonth{February}
\jname{Computer}
\jtitle{Computer}
\pubyear{2026}

\begin{document}

\sptitle{Article Type: Methodology, How to Summarize Numbers Using Means}

\title{One Mean Can't Rule Them All}

\author{Ahmed Nematallah}
\affil{Uppsala University, ahmed.nematallah@it.uu.se}

\markboth{ACCEPTED MANUSCRIPT}{ACCEPTED MANUSCRIPT}

\begin{abstract}\looseness-1There has been extensive debate on how to summarize performance results, especially within the computer architecture community. This article presents a systematic approach to choosing the correct mean, aiming to answer this question for all computer scientists once and for all.
\end{abstract}

\maketitle

\section{Motivation}

\chapteri{F}iguring out how to summarize performance results is challenging, especially given the conflicting literature on the topic. The majority of work presented in MICRO'23 did not state the \whatismean{} used, indicating a lack of clarity in the architecture community. Common misconceptions include using the geometric mean exclusively for normalized values\cite{Fleming86}, and always using the harmonic mean for raw instructions per cycle (IPCs)\cite{Smith86Characterizing, Eeckhout2024RIP}. Additionally, previous work focuses on a specific \whatisipc{} (mainly IPC) instead of providing a general guideline. 
The confusion is worsened by the usage of the means —originally intended to reflect central tendency or expected value— to rank systems by their suitability to run a workload.

\begin{figure}
\includegraphics[width=\linewidth, clip]{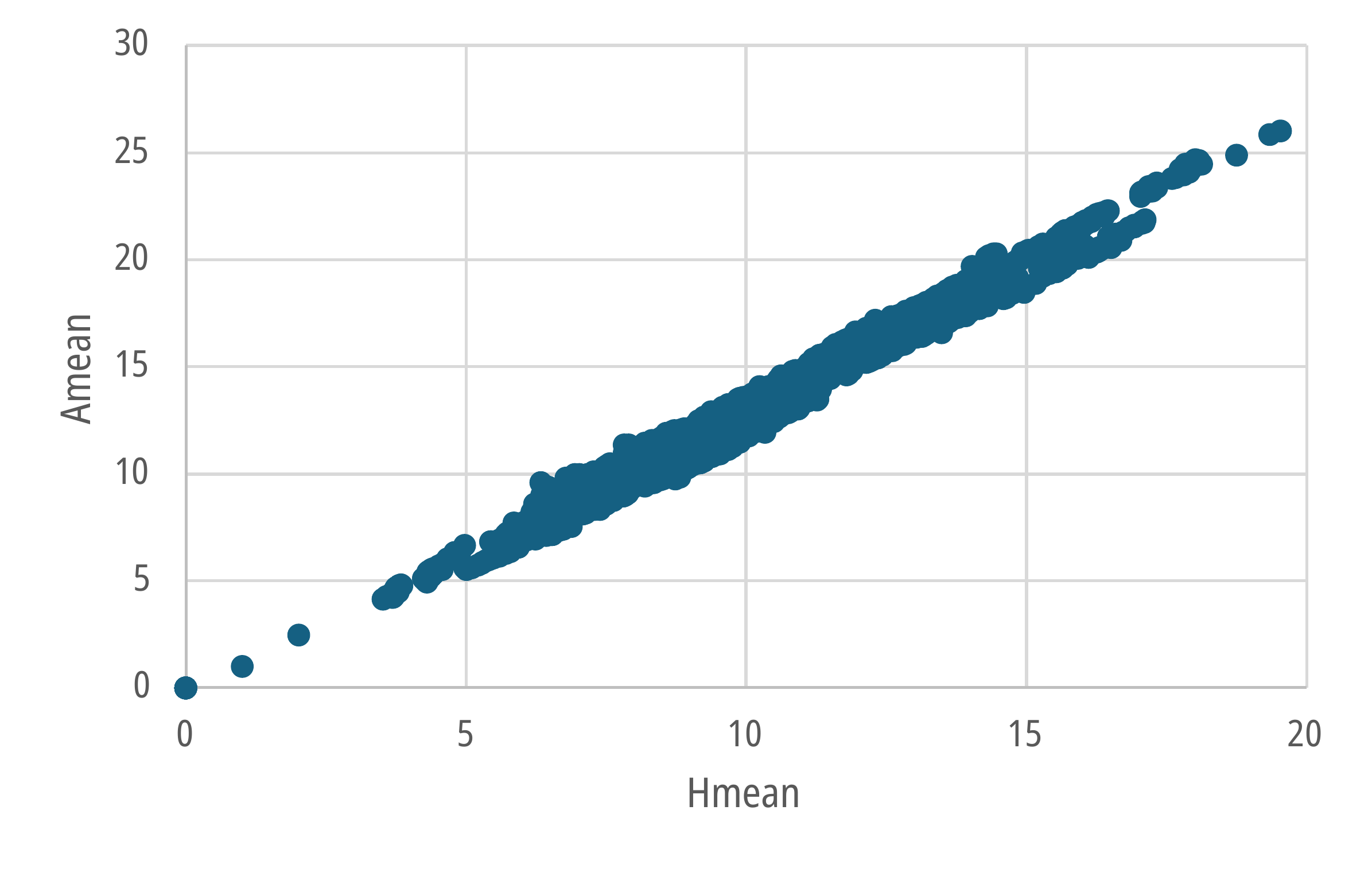}
\includegraphics[width=\linewidth, clip]{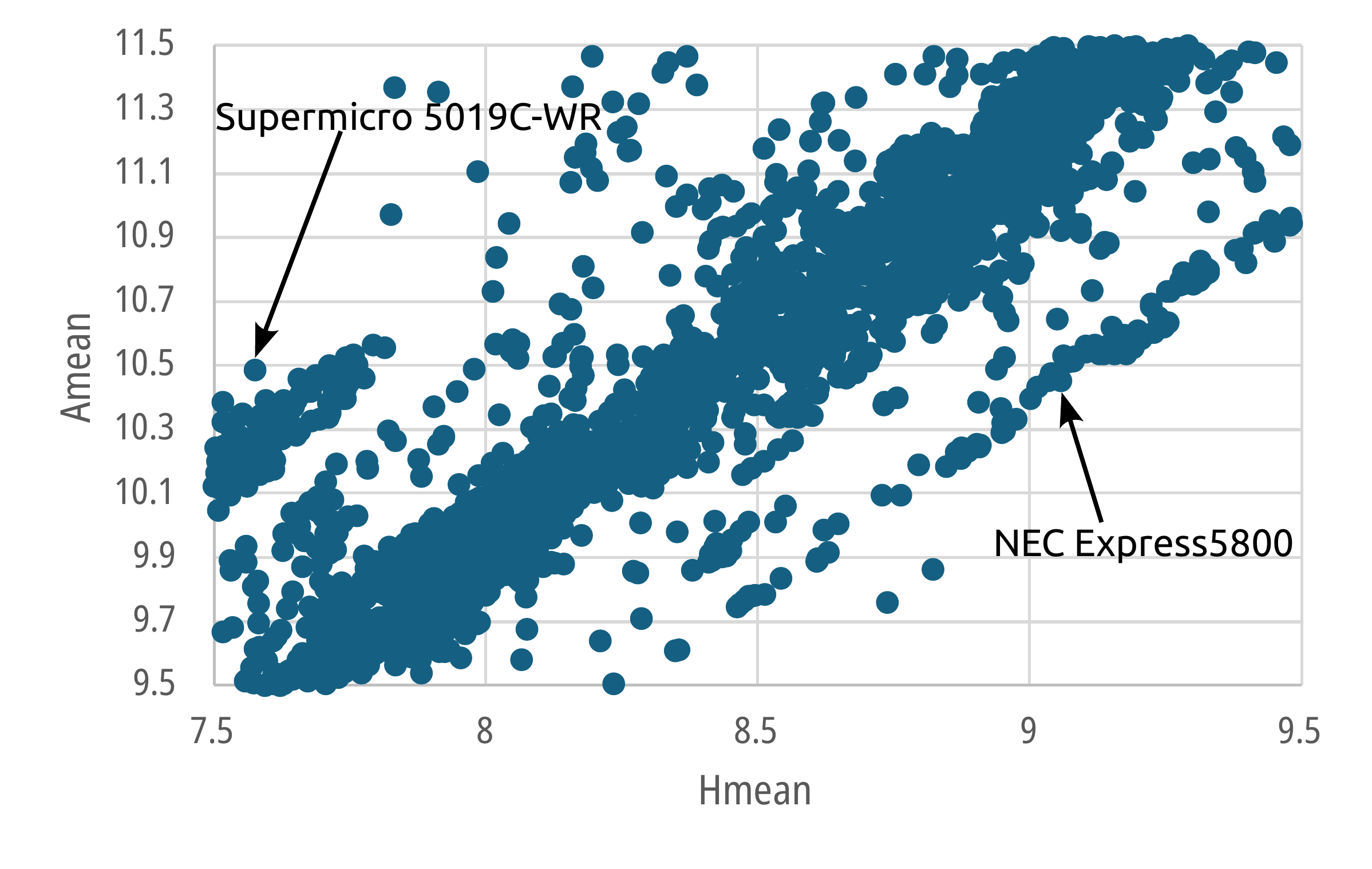}

\caption{
The arithmetic vs harmonic means of SPECRates for over 9000 machines, while there is a strong correlation (top), zooming in (bottom) shows that using the wrong mean can lead to the wrong conclusion, based on\cite{Eeckhout2024RIP}, data from\cite{SPECDB}.
}

\label{fig_diff_geo_harm}
\end{figure}

Some argue that the choice of mean does not matter\cite{Citron2006harmonic}. This is incorrect as using the correct mean ensures that the result is meaningful.  \autoref{fig_diff_geo_harm} further disproves this, as while the harmonic and arithmetic means of SPECrates are correlated, using the wrong mean can be misleading. For instance, the Supermicro 5019C-WR has a slightly higher arithmetic mean than the NEC Express5800, yet the NEC achieves a 19\% higher harmonic mean, which can lead to, for example, buying a less appropriate machine.

\autoref{tab:relatedWorks} compares select prior work to the proposed \whatismeanpl. Some match (\GreenCheck{}), some match for incorrect reasons (\OrangeWarning{}), and many do not match (\RedX{}). The problematic assumptions in previous work can be categorized into: incorrectly assuming the workload is sequential (\textit{Sequential assumption}), focusing on expected values instead of scores (\textit{Expected value}), using an \whatismean{} designed to compare different classes of machines to compare the same class (\textit{Apples to oranges}), assuming the means are all approximately the same (\textit{Triviality assumption}), and finally, assuming that human intuition, rather than solid mathematical foundations are the basis for choosing a mean (\textit{Intuition}). No previous work provides a generic framework for selecting a mean for different applications.

This paper starts by comparing the use of the mean to indicate an expected value vs a score, followed by what means mean and how they work and ends by providing examples of how to use this methodology.

\begin{table*}[t] 
\centering
\caption{Means throughout the years in Computer Architecture. WS: Weighted speedup, FLOPS: FP operations per second.}
\label{tab:relatedWorks}
\begin{tabular}{|l|l|l|l|l|>{\columncolor[rgb]{0.5, 1, 0.5}}l|l|l|}
\hline
Work & Year & Workload     & Metric & Advocates            & This Work & Match & Problem \\ \hline
     SPEC\cite{Spec_metrics} & -    & Benchmarks      & Latency & Geomean            & \GSterm  & \GreenCheck     & -       \\ \hline
     How not to lie...\cite{Fleming86} & 1986 & Benchmarks       & Any & Geomean            & \GSterm  & \GreenCheck     & -       \\ \hline
     Characterizing...\cite{Smith86Characterizing} & 1988 & Benchmarks       & FLOPS & Hmean             & \GSterm  & \RedX     & Sequential assumption \\ \hline
     War of the...\cite{Mashey04War} & 2003 & Benchmarks       & Latency & Geomean            & \GSterm  & \GreenCheck{}\OrangeWarning{}     & Expected value \\ \hline
     The harmonic or...\cite{Citron2006harmonic} & 2006 & Benchmarks       & Any & Any            & \GSterm  & \RedX     & Triviality assumption \\ \hline
     R.I.P geomean...\cite{Eeckhout2024RIP} & 2024 & Benchmarks       & IPC & Hmean             & \GSterm  & \RedX     & Sequential assumption \\ \hline
     Symbiotic...\cite{WSPaper2000} & 2000 & Multiprogram & IPC & WS           & \GSterm  & \RedX     & Apples to oranges \\ \hline
     System-level...\cite{eyerman2008system} & 2008 & Multiprogram & IPC & WS           & \GSterm  & \RedX     & Apples to oranges \\ \hline
     Demystifying...\cite{Michaud2013Demystifying} & 2012 & Multiprogram  & IPC & A/Hmeans & \GSterm  & \RedX     & Assumes intuition \\ \hline
     CRC1...\cite{CRC1} & 2016 & Multiprogram  & IPC & WS           & \GSterm  & \RedX     & Reasoning not stated \\ \hline
     Amdahl's law\cite{amdahl1967validity} & 1967 & End-to-end    & Speedup & Hmean*            & \HSterm   & \GreenCheck{}\OrangeWarning{}    & Not generally applicable \\ \hline
     Simulation Point...\cite{Calder_2005} & 2003 & End-to-end   & IPC & Hmean             & \HSterm   & \GreenCheck     & -       \\ \hline
\end{tabular}
\end{table*}

\section{Expecting a value or a score?} \label{DifferentUsesOfMeans}

``The mean, also called the average, is a measure of central tendency of a group of values."\cite{brillant_mean} It is traditionally used to obtain the expected, or most likely value of a set of data. This is useful when running an experiment multiple times to cancel out error sources. Means here are selected based on the distribution of data; the arithmetic mean should be used if the data are normally distributed, the geometric mean if it is log-normally distributed, and the harmonic mean if the reciprocals of the data are normally distributed. An example of such use case is Ricau's experiments\cite{ricau2023statistically}.

Obtaining a representative value may not always be the goal. Sometimes, the aim is to rank two sets of numbers based on a certain property. In this case, these basic mathematical operations may still be useful. For instance, when evaluating the impact of a disturbance on a baseline: comparing the ``mean" of the baseline numbers to the ``mean" of the numbers after the disturbance, can reveal some properties of this disturbance. 
For example, when summarizing the runtime lengths of different sections of a program, we are not interested in the expected value of each section's runtimes, but rather in which configuration minimizes the total program runtime. Thus, the smallest mean must reflect the configuration with the shortest total runtime. Furthermore, an absolute decrease in runtime for any section must be considered equal as they all contribute equally to decreasing the total runtime.

In this case, the mean should not follow the distribution of the data because we are not interested in an expected value but rather in whether the workload takes longer to run or not.
In fact, in the previous example, if the runtimes are log-normally distributed and the geometric mean is used, the smallest mean may not correspond to the shortest total runtime, rendering the results useless. The selection of the mean should thus follow the properties we are interested in.
To preserve the distinction between both cases, different terms should be used for each one. For the latter case, one suggestion is adding ``As a Score" to the end (i.e. Geometric Mean as a Score (GMS)), or simply calling them scores, (i.e. Geometric Score).

In summary, means are used for two purposes:
\begin{itemize}
    \item To obtain a number representing a group of values. Here, we are interested in an expected value that represents a central, most likely value. An example would be an experiment that is repeated multiple times to remove errors affecting each value.
    The appropriate mean depends on the distribution of data as exemplified by Mashey\cite{Mashey04War} and Iqbal et al\cite{iqbal2010confusion}.
    Here, calling them means is appropriate (e.g. Geometric Mean)
    \item To obtain a score to compare multiple different sets of values. Here, we are interested in which set maximizes (or minimizes) a certain property. Normally, each value in each set would have a corresponding value in the other sets. An example would be different benchmark results across multiple machines.
    The appropriate mean depends on the statistical property of interest. This use case is the focus of this work.
    Adding ``as a Score" to the end (e.g. Geometric Mean as a Score (GMS)) or simply calling them scores (e.g. Geometric Score) helps distinguish both cases.
\end{itemize}

\section{What Do the Means Mean?} \label{MeanReviewSection}

If we care about comparing multiple sets of data instead of obtaining an expected value, the selection of a mean shifts from the data distribution to the \RQtermPl. While we may measure a certain metric (such as IPC), we may be more interested in another metric (such as maximizing throughput). Defining those quantities and their relationship gives the mean a meaning.
For example, does the machine with the highest mean IPC always have the highest throughput?

\subsection{Mathematical Background}

The \ASterm{} is defined as $\AStermshort=\frac{Value_1+Value_2+Value_3...}{n}$. Since addition is associative and commutative, any absolute additive change to any value will affect the mean equally, i.e. $\frac{(Value_1+{\color{red}x})+Value_2+Value_3...}{n}$ equals $\frac{Value_1+(Value_2+{\color{red}x})+Value_3...}{n}$.

The \GSterm{} is defined as $\GStermshort=\sqrt[n]{Value_1*Value_2*Value_3...}$. Multiplication is also associative and commutative, so any multiplicative (percentage/relative) change to any value will affect the mean equally, i.e. $\sqrt[n]{(Value_1*{\color{red}x})*Value_2*Value_3...}$ equals $\sqrt[n]{Value_1*(Value_2*{\color{red}x})*Value_3...}$.

The \HSterm{} is the reciprocal of the \ASterm{} of the reciprocals of the constituent values, and is defined as $\HStermshort=\frac{n}{\frac{1}{Value_1}+\frac{1}{Value_2}+\frac{1}{Value_3}...}$, any absolute change in the reciprocals of the values will affect the mean equally.

Note that since the \GSterm{} is multiplicative, a relative change in any of the reciprocals of the constituent values also affects the mean equally.

\subsection{\texorpdfstring{IPC as an Illustrative Example\protect\footnote{The findings here apply generally to any \RorMQterm. IPC is only used to illustrate the findings.}}{IPC as an Illustrative Example}}\label{sec:consistent_means_for_ipc}

\begin{figure*}[!t]
\centering
\includegraphics[width=\textwidth, trim={0cm 22cm 0cm 14cm},clip]{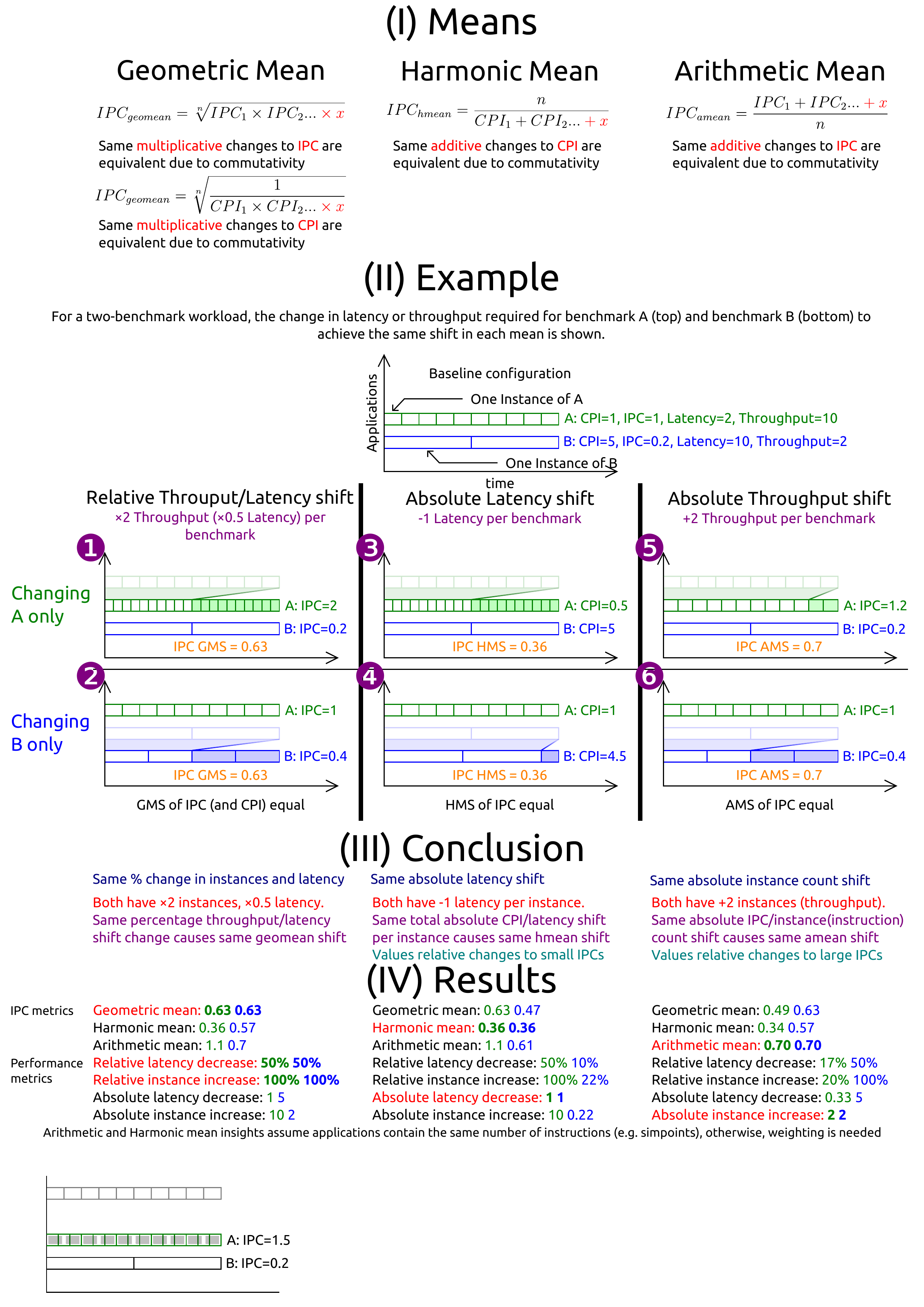}

\caption{
The \RQterm{} (latency/throughput) is changed to achieve the same \MQterm (IPC) mean.
{\color{Green}Benchmark A} and {\color{blue}Benchmark B} are run (sequentially or independently) for 20 time units and each is sped up separately.
}
\label{fig_means_expl}
\end{figure*}

\begin{table}[t]
\caption{Numerical Results for \autoref{fig_means_expl}. TP=Throughput}
\label{tab:fig_numeric}
\setlength\tabcolsep{3pt}
\centering
\begin{tabular}{|l|l|l|l|l|l|l|l|}
\hline & BL & \circleOne{} & \circleTwo{} & \circleThree{} & \circleFour{} & \circleFive{} & \circleSix{} \\ \hline
IPC \GSterm{}         & 0.45 & \cellcolor{Salmon}{\color{ForestGreen} \textbf{0.63}}  & \cellcolor{Salmon}{\color{RoyalBlue} \textbf{0.63}}  & {\color{ForestGreen} 0.63} & {\color{RoyalBlue} 0.47} & {\color{ForestGreen} 0.49} & {\color{RoyalBlue} 0.63} \\ \hline
IPC \HSterm{} & 0.33 & {\color{ForestGreen} 0.36} & {\color{RoyalBlue} 0.57} & \cellcolor{Salmon}{\color{ForestGreen} \textbf{0.36}} & \cellcolor{Salmon}{\color{RoyalBlue} \textbf{0.36}} & {\color{ForestGreen} 0.34} & {\color{RoyalBlue} 0.57} \\ \hline
IPC \ASterm{} & 0.6 & {\color{ForestGreen} 1.1} & {\color{RoyalBlue} 0.7} & {\color{ForestGreen} 1.1} & {\color{RoyalBlue} 0.61} & \cellcolor{Salmon}{\color{ForestGreen} \textbf{0.7}} & \cellcolor{Salmon}{\color{RoyalBlue} \textbf{0.7}} \\ \hline
Rel. Lat. Dec. & 0\%  & \cellcolor{Salmon}{\color{ForestGreen} \textbf{50\%}}  & \cellcolor{Salmon}{\color{RoyalBlue} \textbf{50\%}}  & {\color{ForestGreen} 50\%} & {\color{RoyalBlue} 10\%} & {\color{ForestGreen} 17\%} & {\color{RoyalBlue} 50\%} \\ \hline
Rel. TP Inc. & 0\%  & \cellcolor{Salmon}{\color{ForestGreen} \textbf{100\%}} & \cellcolor{Salmon}{\color{RoyalBlue} \textbf{100\%}} & {\color{ForestGreen} 100\%} & {\color{RoyalBlue} 22\%} & {\color{ForestGreen} 20\%} & {\color{RoyalBlue} 100\%} \\ \hline
Abs. Lat. Dec. & 0 & {\color{ForestGreen} 1} & {\color{RoyalBlue} 5} & \cellcolor{Salmon}{\color{ForestGreen} \textbf{1}} & \cellcolor{Salmon}{\color{RoyalBlue} \textbf{1}} & {\color{ForestGreen} 0.33} & {\color{RoyalBlue} 5} \\ \hline
Abs. TP Inc. & 0 & {\color{ForestGreen} 10} & {\color{RoyalBlue} 2} & {\color{ForestGreen} 10} & {\color{RoyalBlue} 0.22} & \cellcolor{Salmon}{\color{ForestGreen} \textbf{2}}   & \cellcolor{Salmon}{\color{RoyalBlue} \textbf{2}}   \\ \hline
\end{tabular}
\end{table}

When performing experiments, two important metrics arise: a ``\MQterm" and a ``\RQterm", the former refers to the metric being measured, such as the time to complete a workload, IPC, or misses per thousand instructions (MPKI), while the latter refers to a metric we would like to maximize or minimize, such as latency, throughput, or energy consumption. Finding the correct mean relies on finding the relationship between them.

In computer architecture, we are typically interested in minimizing the latency of each application, or maximizing its throughput; thus, both serve as \RQtermPl. IPC is usually easier to measure, especially for microarchitecture research. Thus, it is usually used as a \MQterm. IPC\footnote{CPI, its reciprocal, has the inverse relationship.} is related to both \whatisipcpl{} by $Throughput=\frac{1}{Latency}=\frac{IPC \times Frequency}{InstructionCount}$.

\autoref{fig_means_expl} shows how different IPC means are affected by changes in throughput and latency for a workload composed of two benchmarks: {\color{Green}Benchmark A} (high IPC) and {\color{blue}Benchmark B} (low IPC). Both benchmarks are run for 20 time units either independently or sequentially. We measure the mean IPC when each benchmark is sped up separately (i.e. speeding up {\color{Green}Benchmark A} but not {\color{blue}Benchmark B}, and vice versa). Three speedup cases are shown: (i) a $2\times$ relative IPC increase ($0.5\times$ CPI decrease), (ii) a $-0.5$ absolute CPI decrease, and (iii) a $+0.2$ absolute IPC increase.

In \circleOne{} and \circleTwo{}, the same relative change in throughput or latency (both directly and inversely proportional to the \MQterm, by $2\times$ and $0.5\times$, respectively) yields the same \GSterm{} of IPCs, regardless of which benchmark is affected. This shows that the \GSterm{} is appropriate when relative changes to any value are equally important.

In \circleThree{} and \circleFour{}, an equal absolute change in latency (inversely proportional to the \MQterm, by $-1$ time unit per instance) yields the same \HSterm{} of IPCs, again, independent of which benchmark is affected. This means that applications with a higher baseline IPC (or lower CPI/latency) need to be sped up more in relative terms to achieve the same \HSterm{}. Although this may appear unfair, it can be desirable in some cases as will be discussed later.

In \circleFive{} and \circleSix{}, the same absolute change in throughput (directly proportional to the \MQterm, by $+2$ instances) yields the same \ASterm{}, independent of the affected benchmark. This means that applications with a lower baseline IPC/throughput need to be sped up more in relative terms to achieve the same \ASterm{}, which may also be appropriate in some cases.

Both the \HSterm{} and \ASterm{} IPC insights only hold if the number of instructions is the same for both benchmarks, otherwise, weighting is needed as will be discussed later.

This leads to the following, for any \RandMQterm.
\begin{itemize}
    \item If relative changes to the \RQterm{} matter most use the \GSterm{} on the \MQterm.
    \item If absolute changes to the \RQterm{} matter most:
    \begin{itemize}
        \item Use the \ASterm{} on the \MQterm{} if both are directly proportional.
        \item Use the \HSterm{} on the \MQterm{} if both are inversely proportional.
    \end{itemize}
\end{itemize}

Since the \GSterm{} deals with relative instead of absolute values, it has the special property of providing a qualitative analysis, which is useful when the constituent values do not have any inherent weighting. For example, a benchmark suite, where a relative change to the performance of any constituent benchmark matters equally, regardless of the baseline performance. i.e. applications with high latency are not more or less important than ones with low latency.

\subsection{Weights}
\label{sec:weights}

We discussed how different means preserve different properties, but the insights for the \ASandHSTerm{} do not hold if the instruction count for each benchmark is not the same. This indicates that weighting must be used to preserve the properties for the \RQterm.

For the \ASterm, weighting is done by multiplying each input value by a weight indicating how much it contributes to the final score, then summing up the results: $Amean_{weighted}=w_1 \times value_1+w_2 \times value_2+w_3 \times value_3...$, weighting for the harmonic mean is similar: $\frac{1}{Hmeann_{weighted}}=\frac{w_1}{value_1}+\frac{w_2}{value_2}+\frac{w_3}{value_3}...$
\footnote{The geometric mean is weighted by raising each value to a power then multiplying them, where the sum of all powers is one, i.e. $Gmeann_{weighted}=value_1^{w_1} \times value_2^{w_2} \times value_3^{w_3}...$}.
Note that in both cases, the total sum of all weights must be one.

The weights are taken from the relationship between the \MQterm, and the \RQterm. If the proportionality constant is not the same for all values, weighting will be needed, and this proportionality constant will form the weight. Since the sum of all proportionality constants will not always be one, all weights must be divided by their sum. Weighting can be thought of as partially converting the \MQterm{} to the \RQterm{} then back again.

An example is when the \RQterm{} is latency, the \MQterm{} is IPC, and absolute changes to latency are prioritized. The equation relating IPC and latency is $Latency=\frac{InstructionCount}{IPC \times Frequency}$, here, the frequency is constant throughout all benchmarks, but the number of instructions may not be, so $Latency \propto \frac{InstructionCount}{IPC}$, since the latency is inversely proportional to the IPC, so the \HSterm{} should be used, the weights are the instruction count. Since the sum of all instruction counts may not be equal to one, the instruction count for each benchmark should be normalized to the sum of all instruction counts, i.e. $w_n=\frac{InstructionCount_n}{\sum_{x=1}^NInstructionCount_x}$.

\subsection{Visualizing Means}
\label{sec:visualizing_means}

\begin{figure*}[!t]
\centering
\includegraphics[width=\textwidth, trim={1.5cm 2cm 1.5cm 0cm},clip] {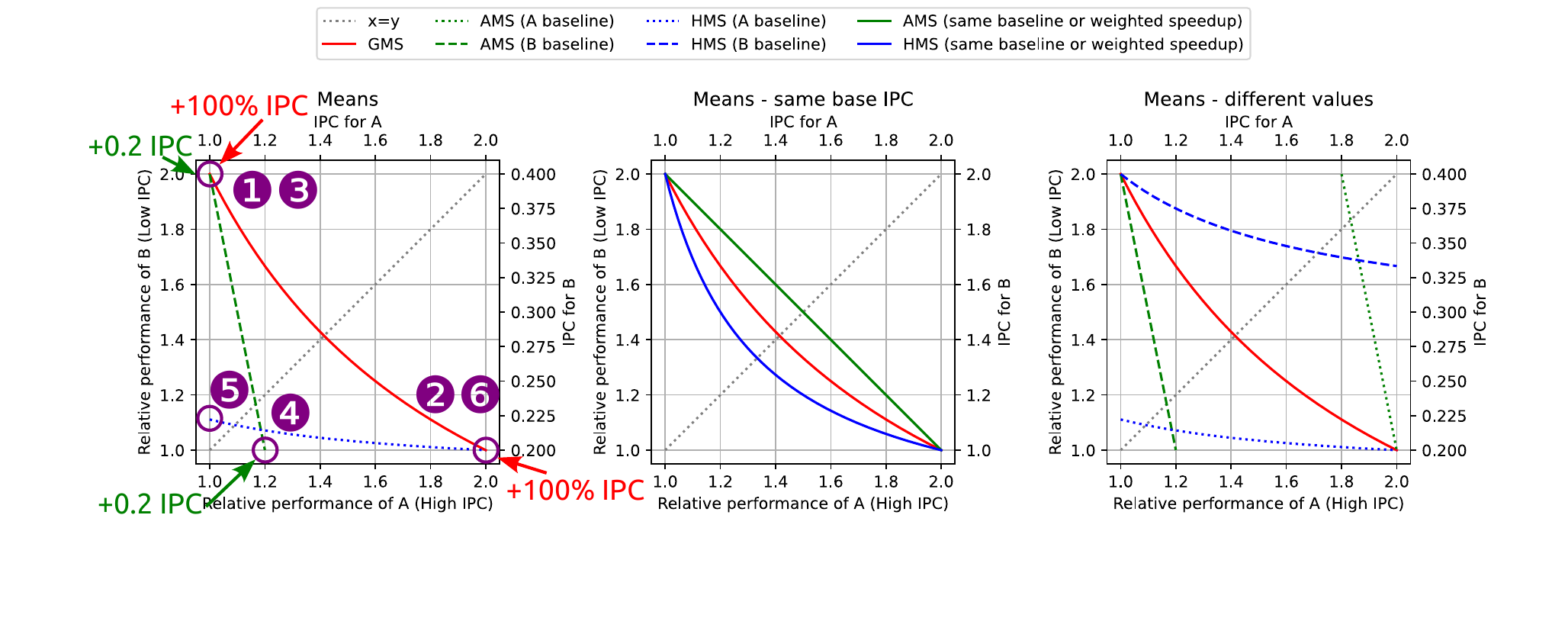}

\caption{
Iso-mean plots (for \autoref{fig_means_expl} benchmarks, {\color{Green}Benchmark A} and {\color{blue}Benchmark B}) depict relative (bottom/left axes) and absolute (top/right axes) speedups. 
Left: The \ASterm{} (green, dashed) holds for an absolute IPC increase of $+0.2$, the \GSterm{} (red, solid) for a $2.0\times$ relative IPC increase, and the \HSterm{} (purple, dotted) for a $-0.5$ CPI decrease.
Middle: Iso-mean lines for applications with equal baseline IPCs or when using weighted speedups. 
Right: Similar to left, includes additional arithmetic ($+1.0$ IPC) and harmonic ($-2.5$ CPI) lines showing how different the results are for reversed relative speedups.
}
\label{fig_means_graph_a}
\end{figure*}

\begin{figure*}[!t]
\centering
\includegraphics[width=\textwidth, trim={1.5cm 1.5cm 1.5cm 0cm},clip] {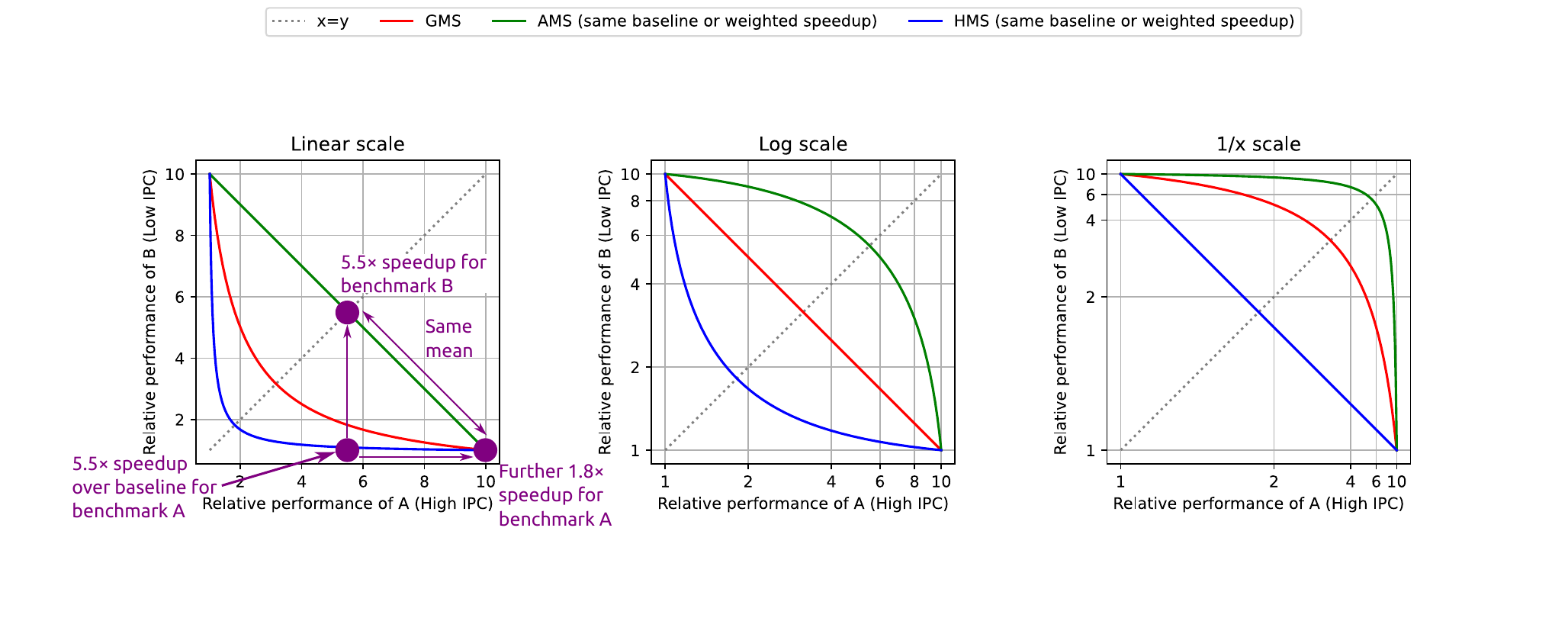}

\caption{
Iso-mean plots following \autoref{fig_means_graph_a} with the maximum speedup increased to $10\times$ to demonstrate how the three means handle outliers differently. Left: linear axes (arithmetic iso-mean is linear), Middle: log axes (geometric iso-mean is linear), Right: $1/x$ axes (harmonic iso-mean is linear). The effects of stacking on the \ASterm{} are shown in purple.
}
\label{fig_means_graph_b}
\end{figure*}

To visualize the difference between the means, the novel "iso-mean line plots" are shown in \autoref{fig_means_graph_a} and \autoref{fig_means_graph_b}.
This plot shows a line where the performance results of two benchmarks (x-axis and y-axis) vary, but combine to produce the same mean.
The benchmarks from \autoref{fig_means_expl} are reused, with the circled numbers indicating the same configurations as before.
The performance of {\color{Green}Benchmark A} and {\color{blue}Benchmark B} is shown on the X-, and Y-Axes, respectively with the relative change (speedup) shown on the main (bottom and left) axes, and the absolute IPC shown on the secondary (top and right) axes.
``A baseline'' and ``B baseline'' indicate the mean that produces a $100\%$ speedup in A with no speedup in B and vice versa, respectively. 

In \autoref{fig_means_graph_a} (Left), the geometric iso-mean line (red) is mirrored across the X=Y line, indicating that a speedup of (a,b) is equivalent to a speedup of (b,a), which means switching the speedup values for two applications produces the same result. The arithmetic iso-mean line preserves absolute changes, as the sum of the IPC shift is the same ($+0.2$ from the baseline).

In \autoref{fig_means_graph_a} (middle), both benchmarks have the same baseline IPC (or are normalized to a baseline as will be discussed later.) removing the asymmetry.
The arithmetic iso-mean line (green) crosses the $x=y$ axis at 1.5, indicating that speeding up one benchmark by $100\%$ produces the same mean as speeding up both by $50\%$.
The geometric iso-mean line (red) equates a 100\% speedup in one to a 40\% speedup in both, while the harmonic iso-mean line (blue) reduces this further to 30\%. This illustrates that the \ASterm{} does not attenuate outliers, whereas the \HSterm{} does so the most.
\autoref{fig_means_graph_b} (left) amplifies the relative increase in IPC from $2\times$ to $10\times$. Here, a $10\times$ speedup in one application is equal to a $5.5\times$, $3.3\times$, and $1.8\times$ speedup in both applications for the \ASandGSandHSTerm, respectively. Showing more clearly how the means handle outliers.

The additive nature of the \ASandHSTerm{} limits the effect of the input values when they approach zero or infinity, respectively.
Assuming all other values remain constant and that there are 10 values, when one value approaches infinity, the \HSterm{} cannot increase by more than $\frac{n_{values}}{n_{values} - 1}$, which is $11\%$.
On the other hand, when a value approaches zero, the \ASterm{} cannot decrease by more than $\frac{1}{n_{values}}$, or $10\%$. The multiplicative nature of the \GSterm{} prevents this issue.

\autoref{fig_means_graph_b} plots the iso-mean lines on linear, log and reciprocal scales showing how each iso-mean line is a straight line in the corresponding scale.

\subsection{Stacked Changes}

When changes are stacked, i.e. when a change to a benchmark is repeated, the mean will still follow the above properties. For example, doubling the IPC of a workload twice results in the same \GSterm{} as doubling the IPC of two different workloads. Similarly, for the \ASterm{}, adding $+1$ twice to a workload's IPC is equivalent to adding $+1$ to the IPC of two different workloads. But adding $+1$ to the same IPC represents a smaller percentage increase the second time, this means the \ASterm{} amplifies repeating relative changes, while the \HSterm{}, similarly, diminishes them.

\autoref{fig_means_graph_b} (left) illustrates this for the \ASterm{}: after a $5.5\times$ speedup for one application, only a further $1.8\times$ speedup is equal to a $5.5\times$ speedup for the other application. The \HSterm{} reverses this; after a $1.8\times$ speedup of one application, a further $5.5\times$ speedup is needed to match a $1.8\times$ speedup for the other application. For the \GSterm{}, equal percentage speedups always yield the same change in mean.

\subsection{Normalizing to a Baseline}
\label{sec:weighted_speedup}

Some studies\cite{WSPaper2000, eyerman2008system, CRC1} propose normalizing values to a baseline before taking the mean —known as weighted speedup— to counter the bias in the arithmetic and harmonic means. However, if the baselines differ for each set of data\cite{shah2022effective}, the results can be misleading or even counter-intuitive. For example, when normalizing the multicore results for each cache replacement policy to the single-core results of the respective policy, the faster policy will score less if it speeds up the system so much that when used in a multicore setup, the system reaches a bottleneck causing the faster system to scale less well than the slower system, even if it was faster in both cases.

When all baselines are the same, this becomes a speedup \whatismean. For the geometric mean, normalization does not affect rankings or relative values because division is commutative, i.e.: $\sqrt{\frac{value^{enhanced}_1}{value^{baseline}_1}\times\frac{value^{enhanced}_2}{value^{baseline}_2}...} = \frac{\sqrt{value^{enhanced}_1\times value^{enhanced}_2}}{\sqrt{value^{baseline}_1\times value^{baseline}_2}}$.

On the other hand, using the arithmetic or harmonic means is problematic because:
\begin{itemize}
    \item The choice of baseline can distort both the relative differences and even rankings\cite{Fleming86}.

    \item While percentage speedups for any application are equal, the non-logarithmic nature of these means causes stacked changes to the same application to have diminishing or growing effects.

    \item The final \whatismeanpl{} are meaningless and are harder to interpret and reason about.
\end{itemize}

This becomes even more convoluted for multiprogrammed workloads, where the ratio is not simply an enhanced system compared to a baseline system, but rather an enhanced system (e.g. with 4MB L3 cache) running in a multicore configuration to a baseline system (e.g. with 1MB L3 cache) running in a single-core configuration\cite{CRC1}, conflating the speedup from the enhancement with the slowdown of sharing resources, along with the complex interactions between them, producing a meaningless \whatismean.

In conclusion, normalizing to a baseline is unnecessary at best and may even be harmful. It can be useful to measure scalability (e.g. of more cores), but should not be used to measure performance.

\subsection{Steps to Choose the Correct Mean}
\label{sec:steps_to_follow}

To select which mean to use in a certain scenario, the following steps should be followed:
\begin{enumerate}
    \item Find out what \whatisipc{} should be maximized or minimized, and whether absolute, or relative changes to it matter more. (\RQterm)
    \item Find out a measurable \whatisipc{} that is related to the \RQterm. (\MQterm)
    \item If relative changes to the \RQterm{} matter most, use the \GSterm{} on the \MQterm.
    \item If absolute changes matter most, figure out the relationship between both \whatisipcpl.
    \begin{itemize}
        \item If the relationship is directly proportional, use the \ASterm{} on the \MQterm.
        \item If the relationship is inversely proportional, use the \HSterm{} on the \MQterm.
    \end{itemize}
    \item If using the \ASterm{} or the \HSterm, figure out the weights using the proportionality constant. The variable components in this proportionality constant should be used as weights, normalized to their sum to make the total equal to one.
\end{enumerate}

\section{The \HSterm} \label{SequentialSection}
The \HSterm{} preserves absolute changes when the \RandMQtermpl{} are inversely proportional.
Examples include Amdahl's law\cite{amdahl1967validity} and SimPoint\cite{Simpoint_paper}.

\subsection{Amdahl's Law on Steroids}
Speeding up a portion of an application has diminishing returns, as the more one portion is sped up, the less significant it becomes to overall performance. Amdahl’s law quantifies this by estimating speedup in a program with one parallelizable and one non-parallelizable section executed sequentially\cite{amdahl1967validity}.

Amdahl's law not only falls in line with this work, but this work also allows us to derive a generalized version allowing arbitrary sections to be sped up independently. Following the above approach, the \RQterm{} is the overall latency of a program, prioritizing absolute changes, the \MQterm{} is the speedup. The runtime of a program section is inversely proportional to speedup, i.e. $new\_time = \frac{original\_time}{speedup}$. The original time serves as the proportionality constant, so weighting should be each section’s fraction of the total original runtime.

\[Speedup_{total}=\frac{1}{\frac{w_1}{s_1} + \frac{w_2}{s_2} + ...}\]

Amdahl's law's workload has a parallel portion $p$ and a serial portion $1-p$, where only $p$ sped up by $s$.

\[Speedup_{total}=\frac{1}{\frac{1-p}{1} + \frac{p}{s}} = \frac{1}{1-p + \frac{p}{s}}\]

Thus, the \HSterm{} is a generalization of Amdahl's law.

\subsection{Sampled Simulation}
\label{sec:sampled_simulation}
Sampled simulation is used to speed up simulation time by only simulating smaller, but  representative, portions (samples) of a benchmark\cite{Simpoint_paper} from which the overall application performance is estimated.
Here, the \RQterm{} is the total execution time with absolute changes being valued equally. When the \MQterm{} is the IPC, the \HSterm{} should be used. Since the proportionality constant is the number of instructions, the weights should be the normalized number of instructions.
Indeed, the authors of SimPoint\cite{Simpoint_paper}, a popular sampled simulation technique, suggest using the weighted arithmetic mean of CPI\cite{Calder_2005}, which is equivalent to the weighted \HSterm{} of IPC. 

\section{The \GSterm} \label{IndependentSection}

Some workloads consist of benchmarks that reflect expected tasks qualitatively but not quantitatively. Benchmark length does not indicate importance\cite{Mashey04War}, so none should be prioritized. Examples include benchmark suites and multiprogrammed workloads.

\subsection{Benchmark Suites}
\label{sec:single_core_benchmark_suites}
SPEC is a benchmark suite comprising independent benchmarks whose length does not reflect their importance\cite{Spec_metrics}.
For example, on the Supermicro 5019C-WR, 600.perlbench\_s takes 503 seconds to execute vs. 113 seconds for 625.x264\_s, but neither is $4.5\times$ more important than the other\cite{SPECDB, Mashey04War}. This means that relative or percentage changes to each benchmark's latency or throughput matter equally and should produce the same mean; therefore, the \GSterm{} must be used.

Although SPEC uses the geometric mean,\footnote{SPEC uses runtime, which is necessary when architectures differ, but is more challenging to obtain than IPC. IPC is suitable when the instructions and frequency are identical.}, it divides each result by that of a reference machine due to the assumption that this specific ratio is log-normally distributed. This assumption is irrelevant, and the \GSterm{} should be used regardless of the distribution since we are not interested in an expected value, but rather in comparing two machines.
(This shows the importance of distinguishing between both use cases.)
Using a reference machine is also superfluous as the multiplicative nature of the geometric mean means that the denominators cancel out when comparing two machines.
On the other hand, this contradicts recent work which suggests using the harmonic mean\cite{Eeckhout2024RIP}, which is only appropriate if the \RQterm{} is absolute reduction in runtime, indicating an implicit assumption that all benchmarks are run back-to-back.

\subsection{Multiprogrammed Workloads}
\label{sec:multiprogrammed_workloads}
Multiprogrammed workloads run benchmark mixes in parallel, restarting any that finish early until all are done in order to keep the pressure on shared resources.
While the benchmarks share resources, they do not communicate or wait for each other. Since they are independent, their baseline performance is not an indicator of their importance, and relative changes to latency and throughput are more relevant, therefore, the \GSterm{} should be used. 
These workloads have primarily used the weighted speedup which yields incorrect results as discussed previously.

\section{The \ASterm} \label{AMeanSection}

The \ASterm{} preserves absolute changes when the \RandMQtermpl{} are directly proportional.

\subsection{MPKI}
MPKI is a common \whatisipc{} used to measure the efficiency of a cache. A higher MPKI means more energy and power are consumed. As a common \MQterm, this is roughly directly proportional to memory power consumption, which is directly proportional to energy used, with the proportionality constant being $time\times IPC$.
When measuring the MPKI for simpoints, the weighted \ASterm{} should be used since the \RQterm{} is an absolute decrease in the energy used, which is directly proportional to the \MQterm{} (MPKI).

On the other hand, for benchmark suites, the MPKI of each benchmark is independent, meaning the \GSterm{} is more appropriate. Most computer architecture work uses the \ASterm{} for benchmark suite MPKI, this can be reasonable assuming all benchmarks are run for an equal amount of time, and have the same IPC as then applications with high MPKI would contribute more to energy consumption than those with low MPKI indicating the importance of clearly stating the underlying assumptions, though misses per thousand cycles (MPKC, $MPKI\times IPC$) may be more suitable since it eliminates the assumption that all IPCs are equal\footnote{This assumption would not affect the usage of the \GSterm{} for IPC as unlike IPC, all misses are valued equally.}.

\subsection{Hit Ratios}
Hit ratios are used to characterize cache performance, if a program is split into multiple sections, and the overall hit ratio is desired, the \RQterm{} is the number of hits, which should be maximized\footnote{Minimizing misses as a \RQterm{} is possible, but weighting is more complex.}, and the \MQterm{} is the hit ratio for each section. As they are directly proportional, the \ASterm{} should be used. The proportionality constant is the number of accesses in each section, which —divided by the total number of accesses— is the weight.

\section{Conclusion} \label{ConclusionSection}
This paper provides a general framework to help summarize performance results by rigorously analyzing the means and their properties.
We make the important distinction between the two use cases for means; as an expected value, and to compare two populations of data.
We call on the research community to distinguish between both use cases by using different names for each use case (e.g. geometric mean vs \GSterm)
This is followed by an analysis of when each mean is most appropriate, along with different aspects of the means such as weights, or normalizing to a baseline.
Finally, we provide examples of applying these properties to reach different means.
What sets this work apart from previous work is mathematically deriving the general properties of each mean and how to use them, instead of using ad-hoc examples and common myths to connect a mean to a certain workload. This work should help guide all computer scientists when summarizing performance.

\section{ACKNOWLEDGMENTS}
This work was supported by the Knut and Alice Wallenberg Foundation through the Wallenberg Academy Fellows Program under Grant 2015.0153, the European Research Council (ERC) under the European Union’s Horizon 2020 Research and Innovation Program under Grant 715283, and the Swedish Research Council under Grant 2019-02429.

\bibliographystyle{IEEEtran}
\bibliography{Ref}

\begin{thebibliography}{10}
\providecommand{\url}[1]{#1}
\csname url@samestyle\endcsname
\providecommand{\newblock}{\relax}
\providecommand{\bibinfo}[2]{#2}
\providecommand{\BIBentrySTDinterwordspacing}{\spaceskip=0pt\relax}
\providecommand{\BIBentryALTinterwordstretchfactor}{4}
\providecommand{\BIBentryALTinterwordspacing}{\spaceskip=\fontdimen2\font plus
\BIBentryALTinterwordstretchfactor\fontdimen3\font minus \fontdimen4\font\relax}
\providecommand{\BIBforeignlanguage}[2]{{%
\expandafter\ifx\csname l@#1\endcsname\relax
\typeout{** WARNING: IEEEtran.bst: No hyphenation pattern has been}%
\typeout{** loaded for the language `#1'. Using the pattern for}%
\typeout{** the default language instead.}%
\else
\language=\csname l@#1\endcsname
\fi
#2}}
\providecommand{\BIBdecl}{\relax}
\BIBdecl

\bibitem{Fleming86}
\BIBentryALTinterwordspacing
P.~J. Fleming and J.~J. Wallace, ``{How not to lie with statistics: the correct way to summarize benchmark results},'' \emph{Commun. ACM}, vol.~29, no.~3, p. 218–221, mar 1986. [Online]. Available: \url{https://doi.org/10.1145/5666.5673}
\BIBentrySTDinterwordspacing

\bibitem{Smith86Characterizing}
\BIBentryALTinterwordspacing
J.~E. Smith, ``{Characterizing computer performance with a single number},'' \emph{Commun. ACM}, vol.~31, no.~10, p. 1202–1206, oct 1988. [Online]. Available: \url{https://doi.org/10.1145/63039.63043}
\BIBentrySTDinterwordspacing

\bibitem{Eeckhout2024RIP}
\BIBentryALTinterwordspacing
L.~Eeckhout, ``{R.I.P}. geomean speedup use equal-work (or equal-time) harmonic mean speedup instead,'' \emph{IEEE Computer Architecture Letters}, vol.~23, no.~1, pp. 78--82, 2024. [Online]. Available: \url{https://doi.org/10.1109/LCA.2024.3361925}
\BIBentrySTDinterwordspacing

\bibitem{Citron2006harmonic}
\BIBentryALTinterwordspacing
D.~Citron, A.~Hurani, and A.~Gnadrey, ``{The harmonic or geometric mean: does it really matter?}'' \emph{SIGARCH Comput. Archit. News}, vol.~34, no.~4, p. 18–25, sep 2006. [Online]. Available: \url{https://doi.org/10.1145/1186736.1186738}
\BIBentrySTDinterwordspacing

\bibitem{SPECDB}
\BIBentryALTinterwordspacing
SPEC, ``{All Published SPEC CPU 2017 Integer Speed Results}.'' [Online]. Available: \url{https://www.spec.org/cpu2017/results/cint2017/}
\BIBentrySTDinterwordspacing

\bibitem{Spec_metrics}
\BIBentryALTinterwordspacing
``O{verview - CPU 2017},'' Nov 2022. [Online]. Available: \url{https://www.spec.org/cpu2017/Docs/overview.html#metrics}
\BIBentrySTDinterwordspacing

\bibitem{Mashey04War}
\BIBentryALTinterwordspacing
J.~R. Mashey, ``{War of the benchmark means: time for a truce},'' \emph{SIGARCH Comput. Archit. News}, vol.~32, no.~4, p. 1–14, sep 2004. [Online]. Available: \url{https://doi.org/10.1145/1040136.1040137}
\BIBentrySTDinterwordspacing

\bibitem{WSPaper2000}
\BIBentryALTinterwordspacing
A.~Snavely and D.~M. Tullsen, ``{Symbiotic jobscheduling for a simultaneous multithreaded processor},'' in \emph{Proceedings of the Ninth International Conference on Architectural Support for Programming Languages and Operating Systems}, ser. ASPLOS IX.\hskip 1em plus 0.5em minus 0.4em\relax New York, NY, USA: Association for Computing Machinery, 2000, p. 234–244. [Online]. Available: \url{https://doi.org/10.1145/378993.379244}
\BIBentrySTDinterwordspacing

\bibitem{eyerman2008system}
S.~Eyerman and L.~Eeckhout, ``{System-level performance metrics for multiprogram workloads},'' \emph{IEEE Micro}, vol.~28, no.~3, pp. 42--53, 2008.

\bibitem{Michaud2013Demystifying}
\BIBentryALTinterwordspacing
P.~Michaud, ``{Demystifying multicore throughput metrics},'' \emph{IEEE Computer Architecture Letters}, vol.~12, no.~2, pp. 63--66, 2013. [Online]. Available: \url{https://doi.org/10.1109/L-CA.2012.25}
\BIBentrySTDinterwordspacing

\bibitem{CRC1}
\BIBentryALTinterwordspacing
``{Cache Replacement Championship}.'' [Online]. Available: \url{https://jilp.org/jwac-1/}
\BIBentrySTDinterwordspacing

\bibitem{amdahl1967validity}
\BIBentryALTinterwordspacing
G.~M. Amdahl, ``{Validity of the single processor approach to achieving large scale computing capabilities},'' in \emph{Proceedings of the April 18-20, 1967, spring joint computer conference}, 1967, pp. 483--485, reprinted in IEEE Solid-State Circuits Society Newsletter, vol. 12, no. 3, pp. 19--20, 2007. [Online]. Available: \url{https://doi.org/10.1109/N-SSC.2007.4785615}
\BIBentrySTDinterwordspacing

\bibitem{Calder_2005}
\BIBentryALTinterwordspacing
B.~Calder, ``{Simulation Point Tutorial and FAQ},'' Jun 2005. [Online]. Available: \url{https://cseweb.ucsd.edu/~calder/simpoint/sim-points-FAQ.htm}
\BIBentrySTDinterwordspacing

\bibitem{brillant_mean}
\BIBentryALTinterwordspacing
Brilliant.org, ``{Mean (Average) | Brilliant Math \& Science Wiki}.'' [Online]. Available: \url{https://brilliant.org/wiki/data-mean/}
\BIBentrySTDinterwordspacing

\bibitem{ricau2023statistically}
\BIBentryALTinterwordspacing
P.-Y. Ricau, ``{Statistically Rigorous Android Macrobenchmarks}.'' [Online]. Available: \url{https://blog.p-y.wtf/statistically-rigorous-android-macrobenchmarks}
\BIBentrySTDinterwordspacing

\bibitem{iqbal2010confusion}
\BIBentryALTinterwordspacing
M.~F. Iqbal and L.~K. John, ``{Confusion by all means},'' in \emph{Proceedings of the 6th International Workshop on Unique chips and Systems (UCAS-6)}, 2010. [Online]. Available: \url{https://lca.ece.utexas.edu/people/iqbal/ucas6-paper.pdf}
\BIBentrySTDinterwordspacing

\bibitem{shah2022effective}
\BIBentryALTinterwordspacing
I.~Shah, A.~Jain, and C.~Lin, ``{Effective mimicry of belady’s min policy},'' in \emph{2022 IEEE International Symposium on High-Performance Computer Architecture (HPCA)}.\hskip 1em plus 0.5em minus 0.4em\relax IEEE, 2022, pp. 558--572. [Online]. Available: \url{https://doi.org/10.1109/HPCA53966.2022.00048}
\BIBentrySTDinterwordspacing

\bibitem{Simpoint_paper}
\BIBentryALTinterwordspacing
T.~Sherwood, S.~Sair, and B.~Calder, ``Phase tracking and prediction,'' in \emph{Proceedings of the 30th Annual International Symposium on Computer Architecture}, ser. ISCA '03.\hskip 1em plus 0.5em minus 0.4em\relax New York, NY, USA: Association for Computing Machinery, 2003, p. 336–349. [Online]. Available: \url{https://doi.org/10.1145/859618.859657}
\BIBentrySTDinterwordspacing

\end{thebibliography}

\begin{IEEEbiography}{Ahmed Nematallah}{\,}is a PhD student at Uppsala University, Sweden. His current research interests include Computer Architecture, Microelectronics, and Security. Contact him at ahmed.nematallah@it.uu.se.\vadjust{\vfill\pagebreak}
\end{IEEEbiography}

\vfill

\end{document}